\input harvmac
\input epsf

\def\CN{{\cal N}}

\Title{\vbox{\baselineskip12pt\hbox{KUCP-0138}\hbox{{\tt hep-th/9906210}}}}
{\vbox{\centerline{Baryon Configurations in the UV and IR Regions}
\vskip12pt\centerline{of Type 0 String Theory}}}

\centerline{Shigenori Seki\footnote{$^\dagger$}{seki@phys.h.kyoto-u.ac.jp} }
\bigskip
{\it \centerline{Graduate School of Human and Environmental Studies}
\centerline{Kyoto University, Kyoto 606-8501, Japan}}

\vskip .3in

\centerline{{\bf abstract}}

The Type 0 string theory is considered as a dual model of a 
non-supersymmetric gauge theory. A background geometry with 
$N$ electric D3-branes is calculated in UV/IR regions. 
In this paper, we study a D5-brane around $N$ D3-branes from the D5-brane
world volume action as in the Type IIB case, and we obtain some baryon 
configurations at UV/IR regions.    

\Date{6/99}

\newsec{Introduction}

From the AdS/CFT correspondence, the Type IIB string theory 
with large $N$ D3-branes is believed 
to be a dual model of the $\CN = 4$ supersymmetric gauge theory, 
and a baryon configuration is realized as a D5-brane wrapped on $S^5$ 
in the Type IIB theory \ref\WGO{E.Witten, 
``Baryons and Branes in Anti de Sitter Space,'' JHEP 9807 (1998) 006, 
{\tt hep-th/9805112}.}\ref\GO{D.Gross and H.Ooguri, 
``Aspects of Large $N$ Gauge Theory Dynamics as seen by String Theory,'' 
Phys.Rev. D58 (1998) 106002, {\tt hep-th/9805129}.}. In 
\ref\CGS{C.G.Callan, A.G\"{u}ijosa and K.G.Savvidy, 
``Baryons and String Creation from the Fivebrane Worldvolume Action,''
{\tt hep-th/9810092}; C.G.Callan, A.Guijosa, K.G.Savvidy and  O.Tafjord, 
``Baryons and Flux Tubes in Confining Gauge Theories from Brane Actions,'' 
{\tt hep-th/9902197}.}, the D5-brane world volume action is studied 
and the static solution is analysed. It has a spike, where the tension 
(the energy per unit radial coordinate distance) 
is equal to that of $N$ fundamental strings\ \ref\CM{
C.G.Callan and J.M.Maldacena, 
``Brane Dynamics From the Born-Infeld Action,'' 
Nucl.Phys. B513 (1998) 198-212, {\tt hep-th/9708147}.}. 
It is a bound state of $N$ quarks, that is a baryon.

Recently a Type 0 string theory has been considered as a dual model of 
a non-supersymmetric gauge theory \ref\KT{I.R.Klebanov and A.A.Tseytlin, 
``D-Branes and Dual Gauge Theories in Type 0 Strings,'' 
Nucl.Phys. B546 (1999) 155-181, {\tt hep-th/9811035}.}.
The Type 0 string theory has two NS-NS and two R-R sectors and 
is classified into two types of models
\ref\JP{J.Polchinski, ``String Theory,'' vol. 2, Cambridge University Press, 1998.}\ ,
$$\eqalign{
{\rm Type\ 0A :}\quad {\rm (NS-, NS-)}\oplus{\rm (NS+, NS+)}
\oplus{\rm (R+, R-)}\oplus{\rm (R-, R+)}\ , \cr
{\rm Type\ 0B :}\quad {\rm (NS-, NS-)}\oplus{\rm (NS+, NS+)}
\oplus{\rm (R+, R+)}\oplus{\rm (R-, R-)}\ . \cr}
$$
The signs $(\pm)$ represent chiralities 
and the ${\rm (NS-, NS-)}$ sector has tachyon ``$T$''. But after tachyon condensation, 
tachyon mass squared is shifted by the five-form field strength.
Since there are two R-R sectors, RR fields are two sets, electric 
and magnetic. 

In the Type 0B theory, the system with large $N$ electric D3-branes 
is studied in 
\ref\KTi{I.R.Klebanov and A.A.Tseytlin, 
``Asymptotic Freedom and Infrared Behavior in the Type 0 String Approach 
to Gauge Theory,'' Nucl.Phys. B547 (1999) 143-156, {\tt hep-th/9812089}.} 
\ref\Mii{J.A.Minahan, 
``Glueball Mass Spectra and Other Issues for Supergravity Duals 
of QCD Models,'' 
JHEP 9901 (1999) 020, {\tt hep-th/9811156}}
\ref\Mi{J.A.Minahan, 
``Asymptotic Freedom and Confinement from Type0 String Theory,'' 
JHEP 9904 (1999) 007, {\tt hep-th/9902074}.}. The metric is set in a form 
\eqn\met{ds^2 = e^{{1 \over 2}\phi}( 
e^{{1 \over 2}\xi - 5 \eta} d\rho^2 + e^{-{1 \over 2}\xi}
(-dt^2 + \sum_{i=1}^3 dx_i dx_i) + e^{{1 \over 2}\xi - \eta}d\Omega_5^2) \ .}
The D3-branes coordinates are $(t, x_1, x_2, x_3)$. 
Then the action is represented as 
\eqn\toda{\eqalign{
S &= \int d\rho \Bigl( {1 \over 2}\dot{\phi}^2 + {1 \over 2}\dot{\xi}^2 
+ {1 \over 4}\dot{T}^2
- 5\dot{\eta}^2 - V(\phi, \xi, \eta, T)\Bigr)\ , \cr
&V(\phi, \xi, \eta, T) = {1 \over 2}T^2e^{{1 \over 2}\phi 
+ {1 \over 2}\xi - 5\eta} + 20e^{-4\eta} - {Q^2 \over f(T)}e^{-2\xi}\ , \cr
&f(T) = 1 + T + {1 \over 2}T^2 \ . \cr
}}
This is a toda like system.

The asymptotic solutions in the UV region 
$(\rho \equiv e^{-y} \ll 1)$ are analysed 
\eqn\uv{\eqalign{
T &= -1 + {8 \over y} + {4 \over y^2}(39 \ln y - 20) + 
O \Bigl({\ln^2 y \over y^3}\Bigr)\ , \cr
\phi &= \ln (2^{15}Q^{-1}) -2 \ln y + {1 \over y}39\ln y + 
O \Bigl({\ln y \over y^2}\Bigr)\ ,\cr
\xi &= \ln (2Q) - y + {1 \over y} + {1 \over 2y^2}(39\ln y - 104) + 
O \Bigl({\ln^2 y \over y^3}\Bigr)\ ,\cr
\eta &= \ln 2 - {1 \over 2}y + {1 \over y} + {1 \over 2y^2}(39\ln y - 38) + 
O \Bigl({\ln^2 y \over y^3}\Bigr)\ ,\cr
}}
where $Q$ is a D3-brane charge.
At the UV limit$(\rho \to 0)$, the metric \met\ becomes 
\eqnn\uvmet
$$\eqalignno{
ds^2 &= 
R_{UV}^2 \Bigl( {dv^2 \over v^2} + {v^2 \over R_{UV}^4}(-dt^2 + 
\sum_{i=1}^3dx_idx_i) + d\Omega_5 \Bigr)\ , &\uvmet \cr
&\phi = \phi_0,\quad R_{UV}^2 
= 2^{-{1 \over 2}}Q^{1 \over 2}e^{{1 \over 2}\phi_0},
\quad v = 2^{-{1 \over 2}}e^{{1 \over 2}\phi_0}\rho^{-{1 \over 4}}\ , & \cr}
$$
and this represents $AdS_5 \times S^5$. 
On the other hand, the asymptotic solutions 
in the IR region $(\rho \equiv e^y \gg 1)$ are evaluated as 
\eqn\ir{\eqalign{
T &= - {16 \over y} - {8 \over y^2}(9 \ln y - 3) + 
O \Bigl({\ln^2 y \over y^3}\Bigr)\ , \cr
\phi &= -{1 \over 2}\ln (2Q^2) + 2 \ln y - {1 \over y}9\ln y + 
O \Bigl({\ln y \over y^2}\Bigr)\ ,\cr
\xi &= {1 \over 2}\ln (2Q^2) + y + {9 \over y} + 
{9 \over 2y^2}(9\ln y - {20 \over 9}) + 
O \Bigl({\ln^2 y \over y^3}\Bigr)\ ,\cr
\eta &= \ln 2 + {1 \over 2}y + {1 \over y} + {1 \over 2y^2}(9\ln y - 2) + 
O \Bigl({\ln^2 y \over y^3}\Bigr)\ .\cr
}}
At the IR limit$(\rho \to \infty)$, the metric \met\ becomes
\eqnn\irmet
$$\eqalignno{
ds^2 &= 
R_{IR}^2 \Bigl( {dv^2 \over v^2} + {v^2 \over R_{IR}^4}(-dt^2 + 
\sum_{i=1}^3 dx_idx_i) + d\Omega_5 \Bigr)\ , &\irmet \cr
&\phi = \phi_\infty,\quad R_{IR}^2 
= 2^{-{3 \over 4}}Q^{1 \over 2}e^{{1 \over 2}\phi_\infty},\quad 
v = 2^{-{1 \over 2}}e^{{1 \over 2}\phi_\infty}\rho^{-{1 \over 4}},& \cr
}$$
and this also describes $AdS_5 \times S^5$ geometry.

In section 2 we study the D5-brane world volume action in  
the $N$ electric D3-brane background in the UV/IR regions 
and analyse the behavior of the tension 
at a spike. 
Section 3 is devoted to some comments about the baryon configuration.

\newsec{The D5-brane world volume action}

We consider the action of a D5-brane wrapped on $S^5$. 
The D5-brane world volume action consists of the Born-Infeld action 
and the WZ term. Since the Born-Infeld action in the Type 0 string 
theory is given in \KT 
\ref\G{M.R.Garousi, ``String Scattering form D-branes in Type 0 Theories,'' 
{\tt hep-th/9901085}.}\ , we can write down D5-brane world volume action 
\eqn\dfa{\eqalign{
S &= - T_5 \int d^6 \sigma h\Bigl({T \over 2}\Bigr) e^{-\phi}
\sqrt{- \det (g_{ab} + F_{ab})} + T_5 \int A_{(1)} \wedge dC_{(4)} \ ,\cr
&h\Bigl({T \over 2}\Bigr) = 1 + {T \over 4} + {3T^2 \over 32} + \cdots \ , \cr}
}
where $g_{ab} (= G_{\mu\nu}\partial_a X^{\mu}\partial_b X^{\nu})$ 
is an induced metric, $F_{ab}(= dA_{(1)})$ is a U(1) gauge field strength 
on the D5-brane and $C_{(4)}$ is a four-form field. 
The $\sigma = (t, \theta, \theta_1, \theta_2, \theta_3, \theta_4)$ is a 
world volume coordinate of the D5-brane embedded on $S^5$. 
For simplicity we consider an SO(5) isometry in the 
$S^5$ and a static solution as \CGS\ . Let $\rho$ and $A_t$ depend only 
on $\theta$ and all other fields be zero. In this setup the action \dfa\ 
becomes 
\eqn\act{
S = T_5 \Omega_4 \int dt d\theta \sin^4 \theta e^{\xi - 2\eta}
\Bigl\{ -h\Bigl({T \over 2}\Bigr)
\sqrt{e^{\phi -5\eta}{\rho'}^2 + e^{\phi - \eta} - {F_{t\theta}}^2}
+ 4A_{t}e^\phi \Bigr\}\ ,}
where $'$ denotes the $\theta$ derivative and $\Omega_4 (= {8\pi^2 \over 3})$ 
is the volume of unit four sphere. 
The gauge field equation of motion is obtained as 
\eqnn\amot
\eqnn\disp
$$\eqalignno{
{\partial \over \partial \theta}K(\theta) 
&= -4 e^{\xi - 2\eta + \phi}\sin^4\theta\ ,&\amot\cr
K(\theta) &\equiv - h\Bigl({T \over 2}\Bigr) \sin^4 \theta e^{\xi - 2\eta}
{\partial_\theta A_t \over \sqrt{e^{\phi - 5\eta} {\rho'}^2 + e^{\phi - \eta} 
-(\partial_\theta A_t)^2}}\ . &\disp\cr
}$$
Partially integrating \act\ , substituting \disp\ into it and changing sign, 
we evaluate the action \act\ in the tree level 
\eqn\ract{
E_\rho = T_5 \Omega_4 \int d\theta 
\sqrt{e^{\phi - 5\eta}{\rho'}^2 + e^{\phi - \eta}}
\sqrt{K(\theta)^2 + h^2 \sin^8 \theta e^{2\xi - 4\eta}}\ .
}
This depends only on the $\rho$ field.

\subsec{The UV region}

At the UV limit $(\rho \to 0)$ Eq.\ract\ is expressed by using Eq.\uvmet\ as 
\eqn\uvactlim{
T_5 \Omega_4 \int d\theta \sqrt{{v'}^2 + v^2}
\sqrt{K(\theta)^2 + {Q^2 \over 4} 
h\Bigl(-{1 \over 2}\Bigr)^2 \sin^8 \theta }\ ,
}
and Eq.\amot\ is written as 
\eqn\uvdisplim{
\partial_\theta K(\theta) = -4 {R_{UV}}^4 \sin^4 \theta\ .
}
Since the right hand side of \uvdisplim\ is a function of only $\theta$ 
not $\rho$, $K(\theta)$ is solvable and is determined as 
$$\eqalign{
K(\theta) &= {R_{UV}}^4 k(\theta)\ ,\cr
&k(\theta) \equiv {3 \over 2}(\nu \pi - \theta) + 
{3 \over 2}\sin \theta \cos \theta + \sin^3 \theta \cos \theta \ ,\cr }
$$
where $0 < \nu < 1$ is an integral constant. These results are similar to 
those in the Type IIB case \CGS\ , where $\nu = 0$ corresponds 
to the baryon vertex and the spike sticking out at $\theta = \pi$ has the 
tension equal to that of the $N$ fundamental strings. 
So in this paper mainly we concentrate on  
the tension at $\nu = 0$ and $\theta = \pi$. Then Eq.\uvactlim\ implies that 
the tension at $\theta = \pi$ is 
$T_5 \Omega_4|K(\pi)| = T_5 \Omega_4 {R_{UV}}^4{3\pi \over 2}$. 

Next we consider in the near UV region $(\rho \ll 1)$. 
From Eqs.\uv\ and \amot\ , the behavior of the $K(\theta)$ is given as 
\eqnn\uvk
$$\eqalignno{
K(\theta) &= e^{\xi - 2\eta + \phi} k(\theta)\ , &\cr
&\approx \Bigl(1 - {1 \over y} - {1 \over 2y^2}(39 \ln y + 27) + \cdots\Bigr)
{Qg_s \over 2} k(\theta)\ , &\uvk\cr
}$$
where $g_s = e^\phi$ is the string coupling. Substituting \uv\ into \ract\ ,
we can evaluate the $E_\rho$, 
$$\eqalign{
E_\rho = &T_5 \Omega_4 \int d\theta 
\sqrt{e^{{39 \ln y -5 \over y}-{5(39\ln y -38)\over 2y^2} + \cdots}v'^2 + 
e^{{39\ln y -1 \over y}-{39\ln y-38 \over 2y^2} + \cdots} v^2} \cr
&\times \sqrt{\Bigl\{\Bigl(1 - {1 \over y} 
- {1 \over 2y^2}(39 \ln y + 27) + \cdots\Bigr) 
{Qg_s \over 2}k(\theta)\Bigr\}^2 + e^{2\xi -4\eta}h^2 \sin^8 \theta}\ .\cr
}$$
These imply that the tension of the D5-brane at $\theta = \pi$ 
and $\nu = 0$ is expressed as\foot{At the spike, $v'$ will dominate $v$.} 
\eqnn\uvten
$$\eqalignno{
&T_5 \Omega_4 \Bigl(1 + {1 \over 2y}(39\ln y -7) + 
{1 \over 8y^2}(1521(\ln y)^2 - 1092 \ln y +317) +\cdots \Bigr)
{Qg_s \over 2}|k(\pi)| &\cr
=& p(\rho)T_5 \Omega_4{Qg_s \over 2}|k(\pi)| \ ,&\uvten\cr
&p(\rho) \equiv 1 - {39\ln (-\ln \rho) -7 \over 2\ln \rho} 
+ {1521(\ln (- \ln \rho))^2 - 1092 \ln (-\ln \rho) +317 
\over 8(\ln \rho)^2}+\cdots \ . &\cr
}$$

Note that this tension includes a coupling constant $g_s$ 
and seems almost zero because the leading effective gravity solution \uv\ is 
asymptotically free in the UV region,  
$$g_s = e^{\phi} =2^{15}Q^{-1}(-\ln \rho)^{-2-{39 \over \ln \rho} + \cdots}\ .$$
But it is not a problem. The D5-brane's tension \KT\ , 
\eqn\ften{
T_5 = {1 \over \sqrt{2}(2\pi)^5(l_s)^6 g_s}\ ,
} 
where $l_s$ is the string length, has also an inverse coupling. 
So effects of the couplings are canceled and 
the tension does not become almost zero in the UV region. 
The behavior of \uvten\ is shown in fig.\ 1 .

\bigskip
\vbox{
\centerline{\epsfbox{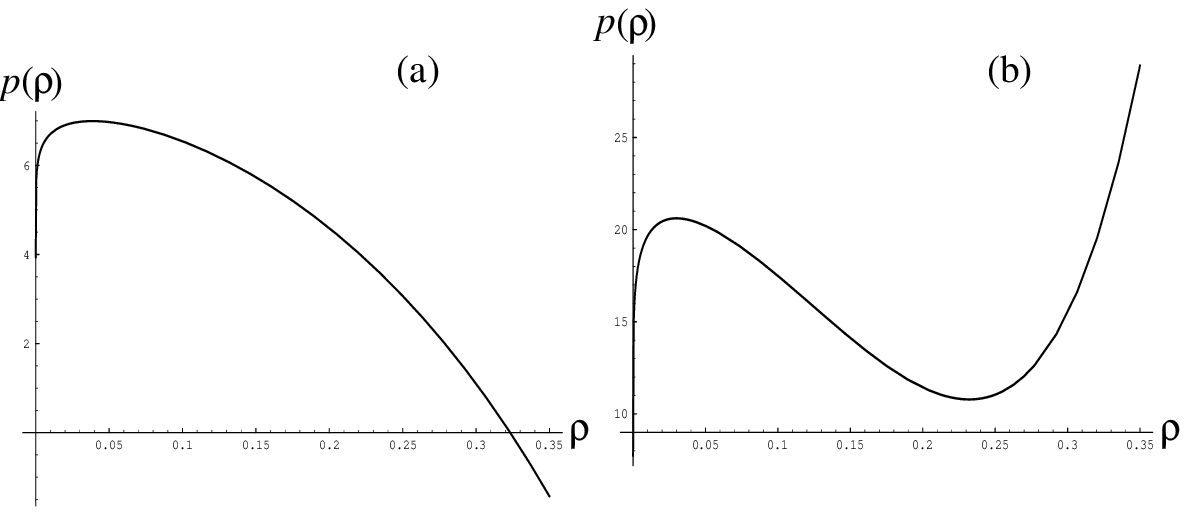}}
\vskip .5cm
\centerline{\fig\figii{The behavior of $p(\rho)$ in the UV region 
$\rho \ll 1$ . (a) The leading order approximation of $p(\rho)$, 
that is, $1 - {39\ln (-\ln \rho) -7 \over 2\ln \rho}$.  
(b) The second order approximation of $p(\rho)$ ,
that is, $1 - {39\ln (-\ln \rho) -7 \over 2\ln \rho} 
+ {1521(\ln (- \ln \rho))^2 - 1092 \ln (-\ln \rho) +317 
\over 8(\ln \rho)^2}$.}\ 
The behavior of $p(\rho)$ in the UV region $\rho \ll 1$ .
}}
\bigskip

In \figii, the (a) is  the leading order approximation of $p(\rho)$, 
that is, $1 - {39\ln (-\ln \rho) -7 \over 2\ln \rho}$ and the  
(b) represents the second order approximation of $p(\rho)$ ,
that is, $1 - {39\ln (-\ln \rho) -7 \over 2\ln \rho} 
+ {1521(\ln (- \ln \rho))^2 - 1092 \ln (-\ln \rho) +317 
\over 8(\ln \rho)^2}$. 
In the Type IIB case, the tension is constant since the radii of $S^5$ and 
$AdS_5$ are 
independent of $\rho$. On the other hand in the Type 0 case, the tension is not
constant and flows as shown in \figii\ \ since the radii of $S^5$ and $AdS_5$ 
depend on $\rho$ \KTi\ . The (a) implies that there exist two points 
$\rho = 0$, and $\rho = \exp(-\exp({7 \over 39})) \approx 0.302219$, 
where $p(\rho)=1$. 
The latter point is an interesting and 
non trivial point, but it is less credible since in this section the 
$\rho \ll 1$ region is analyzed. In fact this point vanishes 
in the second order approximation in the (b). From these analyses, we 
conclude that the tension increases away from the UV limit.
We will give more comments in section 3. 

\subsec{The IR region}

Since at the IR limit the metric \irmet\ describes 
$AdS_5 \times S^5$ geometry as in the
UV limit, the similar behavior will occur also in the IR region.

At the IR limit $(\rho \to \infty)$ the action \act\ is calculated 
by Eq.\irmet\ 
\eqn\iractlim{
\int d\theta \sqrt{{v'}^2 + v^2}\sqrt{K(\theta)^2 + {Q^2 \over 8} 
h(0)^2 \sin^8 \theta}\ ,}
and Eq.\amot\ is expressed as 
\eqn\irdisplim{
\partial_\theta K(\theta) = -4{R_{IR}}^4 \sin^4\theta\ .
}
In the same way as the UV case, we concentrate on the tension at $\theta = \pi$
with $\nu = 0$. Then Eq.\iractlim\ implies 
that the tension at the spike is $T_5 \Omega_4 {R_{IR}}^4 {3\pi \over 2}$. 
Since the $R_{IR}, R_{UV}$ have the following expressions, 
\eqn\rel{
{R_{UV}}^4 = {Q \over 2}g_s, \quad {R_{IR}}^4 = {Q \over 2\sqrt{2}}g_s}
the ratio of tensions at the UV and the IR limits is tured out to be 
$1 \over \sqrt{2}$.

Next we consider at the near IR region $(\rho \gg 1)$ . 
Substituting \ir\ into \amot\ , we obtain the asymptotic form of the 
$K(\theta)$
$$
K(\theta) \approx 
\Bigl( 1 + {7 \over y} + {1 \over 2y^2}(63 \ln y + 33) + \cdots \Bigr)
{Qg_s \over 2\sqrt{2}}k(\theta)\ .
$$
Substituting \ir\ into \ract\ , we calculate the $E_\rho$
$$\eqalign{
E_\rho = &T_5 \Omega_4 \int d\theta 
\sqrt{e^{-{9\ln y + 5 \over y} - {5(9\ln y -2) \over 2y^2} + \cdots} v'^2 + 
e^{-{9\ln y +1 \over y} - {9\ln y -2 \over 2y^2} + \cdots}v^2}  \cr
&\times \sqrt{\Bigl\{\Bigl(1 + {7 \over y} 
+ {1 \over 2y^2}(63 \ln y + 33) + \cdots\Bigr)
{Qg_s \over 2\sqrt{2}}k(\theta)\Bigr\}^2 + e^{2\xi -4\eta}h^2 \sin^8\theta}\ .
}$$
From these equations the tension at $\theta = \pi$ is obtained as
\eqnn\irten
$$\eqalignno{
&T_5 \Omega_4 \Bigl(1 - {1 \over 2y}(9 \ln y - 9) 
+ {1 \over 8y^2}(81(\ln y)^2 + 37) + \cdots \Bigr)
{Qg_s \over 2\sqrt{2}}|k(\pi)| &\cr
=& q(\rho) T_5 \Omega_4 {Qg_s \over 2 \sqrt{2}}|k(\pi)| \ , &\irten \cr
&q(\rho) \equiv 1 - {9 \ln \ln \rho -9 \over 2\ln \rho} 
+ {81 (\ln \ln \rho)^2 + 37 \over 8(\ln \rho)^2} + \cdots \ . &\cr
}$$

Note that this tension seems to diverge apparently since the string coupling 
$g_s$ behaves
$$
g_s = e^\phi = 2^{-{1 \over 2}}Q^{-1}
(\ln \rho)^{2 -{9 \over \ln \rho} + \cdots}\ .
$$
But for the same reason as the UV case, the coupling $g_s$ is canceled 
with the one in the tension \ften\ . 
The behavior of \irten\ is shown in fig.\ 2. 

\bigskip
\vbox{
\centerline{\epsfbox{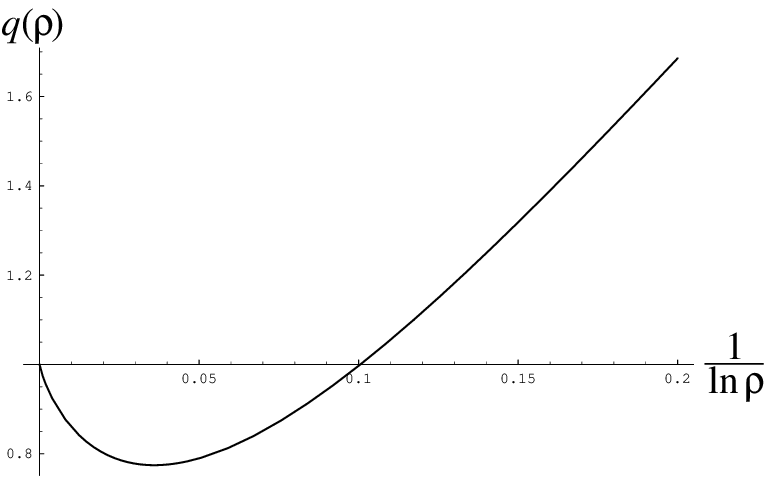}}
\vskip .5cm
\centerline{\fig\figiii{The behavior of $q(\rho)$ in the IR region $(\rho \gg 1)$ .}\ 
The behavior of $q(\rho)$ in the IR region $(\rho \gg 1)$ .}}
\bigskip

In \figiii\ there are also two points $\rho = \infty$ and numerically 
$\rho = 2.08581 \times 10^4$, where $q(\rho)=1$. 
This is the same situation as the UV case. 
Note that the second point is not reliable, because the IR  region has a
very large coupling and the perturbative analysis is no longer credible. 

\newsec{Discussion}

In this paper we studied the static solution of D5-brane worldvolume action
in the UV/IR regions in the background with large $N$ electric D3-branes. 

In the Type IIB case a relation $R^4 = 4\pi g_s N {l_s}^4$ is satisfied, 
where $R$ is the radius 
of $AdS_5$ and $S^5$ and the $N$ is the number of D3-branes. 
This number comes from the action 
$\int d^{10} x \sqrt{G}(\cdots -{1 \over 5!}{F_5}^2)$ 
associated with the five form field strength. 
On the other hand, at the UV limit of the Type 0B, a relation
${R_{UV}}^4 = {Q \over 2}g_s$ is obtained from Eq.\uvmet\ 
up to normalization and the five form field strength appears in the 
action \KT\ , in a form,  
$$
\int d^{10}x \sqrt{G}\Bigl(\cdots - f(T){1 \over 5!}{F_5}^2\Bigr)\ .
$$
This reads that  $f(T)^{-{1 \over 2}}$ contributes to $N$ in the Type 0 case
 ({\it cf.} \toda\ ). 
So using the same normalization as the type IIB case, we can calculate the 
${R_{UV}}^4$ as, 
\eqn\uvre{
{R_{UV}}^4 = 4 \pi g_s {N \over f(-1)^{1 \over 2}} {l_s}^4 
= 4\sqrt{2} \pi g_s N {l_s}^4\ . 
}
In the same way for the IR limit, we obtain 
\eqn\irre{
{R_{IR}}^4 = 4 \pi g_s {N \over f(0)^{1 \over 2}} {l_s}^4 
= 4 \pi g_s N {l_s}^4\ .
}
Eqs.\uvre\ and \irre\  are consistent with the \rel\ .
Then from Eqs.\ften\ and \uvre\ , the tension becomes 
$$
T_5 \Omega_4 {R_{UV}}^4 |k(\pi)| = {N \over 2 \pi {l_s}^2} = N T_f\ ,
$$
where $T_f$ is a tension of a fundamental string. This 
corresponds to 
the bound state of $N$ quarks in SU($N$) gauge theory, i.e. a baryon. 
As mentioned in section 2.1, in the UV region, there exist two points, 
the UV limit point $\rho = 0$ and the non trivial one $(\rho = 0.268067)$, 
where the tension is equal to $NT_f$. But the latter is less credible, because 
Eq.\uvten\ is reliable only in small $\rho$ region and in fact the second 
point disappears in the analysis including the second order 
approximation. It would be interpreted as some perturbative effects. 

Since in the IR limit, from Eqs.\ften\ and \irre\ , we obtain the formula 
of tension, 
$$
T_5 \Omega_4 {R_{IR}}^4 |k(\pi)| = {N \over 2\sqrt{2} \pi {l_s}^2} 
= {N \over \sqrt{2}} T_f\ .
$$
In this case, the number of fundamental strings is 
not integer and this state is not a 
standard baryon. It may represent some quantum effects in such a confinement 
region with a strong coupling. 

If the flow of tension is connected smoothly 
from the UV limit to the IR limit, 
there must exist at least one point where a baryon configuration 
with $N$ quarks is realised, since the flow increases near the UV limit.
But we are not able to determine this interesting point exactly, since 
the full metric of D3-brane background has not been known yet and we cannot discuss 
how the UV solution is connected with the IR solution. 
If the full metric is found, the behavior of that non trivial point 
will be clarified.

Though in this paper the background with only the $N$ electric D3-branes 
is discussed, 
this method will be applicable to more general backgrounds, for example, a
model with  
$N$ electric and $N$ magnetic D3-branes, in which case the dual 
theory is an SU($N$)$\times$SU($N$) Yang-Mills theory 
\ref\KTii{I.R.Klebanov and A.A.Tseytlin, 
``A Non-supersymmetric Large N CFT from Type 0 String Theory,'' 
JHEP 9903 (1999) 015, {\tt hep-th/9901101}.} 
\ref\TZ{A.A.Tseytlin and K.Zarembo, 
``Effective Potential in Non-supersymmetric $SU(N) \times SU(N)$ 
Gauge Theory and Interactions of Type 0 D3-branes,'' {\tt hep-th/9902095}.}\ .

\bigbreak\bigskip\bigskip\centerline{{\bf Acknowledgement}}\nobreak

I am grateful to K.Sugiyama for useful discussions. This work is 
supported in part by JSPS Research Fellowships for Young Scientists (\#4783).

\listrefs
%\listfigs
\bye